\documentstyle[preprint,aps]{revtex}
\begin{document}
\preprint{UM-P-93/58, OZ-93/13}
\draft
\title{Anomalous $WWZ$ couplings and
$ K_L \rightarrow \mu^+\mu^-$}
\author{Xiao-Gang He}
\address{Research Center for High Energy Physics\\
School of Physics\\
University of Melbourne \\
Parkville, Vic. 3052 Australia}
\date{June, 1993 (Revised July, 1993)}
\maketitle
\begin{abstract}
We study contributions to $K_L \rightarrow \mu^+\mu^-$ from anomalous $WWZ$
interactions. There are, in general, seven anomalous couplings.
Among the seven anomalous couplings,
only two of them contribute significantly. The others
are suppressed by factors like $m_s^2/M_W^2$, $m_d^2/M_W^2$, or $m_K^2/M_W^2$.
Using the experimental data on $K_L\rightarrow \mu^+\mu^-$,
we obtain strong bounds on the two  anomalous couplings.
\end{abstract}
\newpage
In this paper we study contributions to $K_L\rightarrow \mu^+\mu^-$ from the
anomalous $WWZ$ interactions.
The Minimal Standard Model of electroweak interactions is in very good
agreement with present experimental data. However its structure should be
tested in detail in order to finally establish the model. One of the
important aspects is to test the structure of self-interactions of electroweak
bosons. Such test will provide information about
 whether the weak bosons are gauge particles with interactions
predicted by the MSM, or gauge particles of
some extensions of the MSM which predict different
interactions at loop levels, or even non-gauge particles whose
self-interactions at low energies are described by effective interactions.
In general there will be more self-interaction terms than the tree level
MSM terms (the anomalous couplings)\cite{anom}.
It is important to find out experimentally what
are the allowed regions for these anomalous couplings.
The process $K_L\rightarrow \mu^+\mu^-$ has been studied in the MSM
extensively\cite{kmu}. It has been used to study the allowed range for
the top quark mass
and the allowed ranges for some of the KM matrix elements. In this paper we
show that $K_L\rightarrow \mu^+\mu^-$ also puts very strong constraints on
some of the $WWZ$ anomalous couplings.

The most general form for the anomalous $WWZ$
interactions can be parametrized as
\begin{eqnarray}
L &=&-gcos\theta_W\Big[ig_1^Z(W^{+\mu\nu}W^{-\mu}Z^\nu - W^+_\mu W^{-\mu\nu}
Z_\nu)\nonumber\\
&+&i\kappa^Z W^+_\mu W^-_\nu Z^{\mu\nu}
+ i {\lambda^Z\over M_W^2}
W^+_{\sigma\rho}W^{-\rho\delta} Z^\sigma_\delta\nonumber\\
&+&  i\tilde \kappa^Z W^+_\mu W^-_\nu \tilde Z_{\mu\nu}
+ i{\tilde\lambda^Z \over M_W^2}W^+_{\sigma\rho}W^{-\rho\delta}\tilde
Z^\sigma_\delta\nonumber\\
&+&g^Z_4W^+_\mu W^-_\nu (\partial^\mu Z^\nu + \partial^\nu Z^\mu)\nonumber\\
&+&g^Z_5\epsilon_{\mu\nu\alpha\beta}(W^{+\mu}\partial^\alpha W^{-\nu}
- \partial^\alpha W^{+\mu} W^{-\nu})Z^\beta\Big] \;,
\end{eqnarray}
where $W^\pm_\mu$ and $Z_\mu$ are the W-boson and Z-boson fields,
$W_{\mu\nu}$ and $Z_{\mu\nu}$ are
the W-boson and Z-boson field strengths, respectively; and $\tilde Z_{\mu\nu}
={1\over 2}\epsilon_{\mu\nu\alpha\beta}Z^{\alpha\beta}$.
The terms proportional to $g_1^Z$, $\kappa^Z$,  $\lambda^Z$ and $g_5^Z$
are CP conserving and $\tilde
\kappa^Z$, $\tilde\lambda$ and $g_4^Z$ are CP violating.

To obtain amplitude for the process $K_L\rightarrow \mu^+\mu^-$, we first
evaluate the effective coupling for $d s Z$ with the Z-boson
off-shell. This coupling
is induced at the one loop level. The
effective Hamiltonian is given by
\begin{eqnarray}
H_{eff} &=& -igcos\theta_W{g^2\over 2}\epsilon^{Z\mu} V_{ld}V_{ls}^* \bar d
\gamma_\alpha \gamma_\nu \gamma_\beta {1-\gamma_5 \over 2} s\nonumber\\
&\times&\int {d^4k\over (2\pi)^4} {k^\nu (g^{\alpha\alpha'} -
{k^{+\alpha} k^{+\alpha'}\over M_W^2})
(g^{\beta\beta'}-{k^{-\beta}k^{-\beta'}\over M_W^2})\Gamma_
{\mu\alpha'\beta'}(q, k^+, k^-)
\over (k^2-m_l^2)((p-k)^2-M_W^2)
((p'-k)^2-M_W^2)} + H.C.\;,
\end{eqnarray}
where $l$ is summed over $u\;, s\;$, and $t$, and
\begin{eqnarray}
\Gamma_{\mu\alpha\beta}(q, k^+, k^-)&=&
g_1^Z(g_{\alpha\beta}(k^-_\mu - k^+_\mu)
+g_{\beta\mu}k^+_\alpha - g_{\alpha\mu}k^-_\beta)\nonumber\\
&-&\kappa^Z (g_{\alpha\mu}q_{\beta}-g_{\beta\mu}q_{\alpha})
-\tilde \kappa^Z \epsilon_{\mu\alpha\beta\rho}q^\rho \nonumber\\
&+& {\lambda^Z\over M_W^2}(g_\alpha^\rho k^{+\delta} - g_\alpha^\delta
k^{+\rho}) (g_{\beta\delta} k_{-\sigma} - g_{\beta\sigma} k_{-\delta})
(g_{\rho\mu} q^\sigma - g^\sigma_\mu q_\rho)\nonumber\\
&+&{\tilde \lambda^Z\over M_W^2}(g_\alpha^\rho k^{+\delta} - g_\alpha^\delta
k^{+\rho})(g_{\beta\delta} k^{-\sigma} -g_\beta^\sigma k_{-\delta})
\epsilon_{\rho\sigma\mu\tau}q^\tau\nonumber\\
&+&ig_4^Z(g_{\beta\mu}q_\alpha+g_{\alpha\mu}q_\beta) +
ig_5^Z\epsilon_{\mu\alpha\beta\sigma}(k^{+\sigma} - k^{-\sigma});,\nonumber
\end{eqnarray}
where $k$, $p$, and $p'$ are the internal,
s-quark and d-quark momenta respectively,
$q = p' - p$, $k^+ = p - k$ and $k^- = k -p'$, and $\epsilon^{Z\mu}$
is the Z-boson
polarization vector. Performing the standard Feynman parametrization, we have
\begin{eqnarray}
H_{eff} &=&- ig^3cos\theta_W\epsilon^{Z\mu} V_{ls}V_{ld}^*\bar  d
\gamma_\alpha\gamma_\nu\gamma_\beta {1-\gamma_5\over 2}s\nonumber\\
&\times& \int^1_0dx \int^{1-x}_0dy \int {d^4k\over (2\pi)^4}
{k^\nu(g^{\alpha\alpha'} - {k^{+\alpha}k^{+\alpha'}\over M_W^2})
(g^{\beta\beta'} - {k^{-\beta}k^{-\beta'}\over M_W^2})\Gamma_{\mu\alpha'\beta'}
(q,k^+,k^-)\over
(k^2 - 2k\cdot(xp+yp')-(m_l^2 +(M_W^2 -m_l^2)(x+y))^3}\;\nonumber\\
&+&H.C.\;.
\end{eqnarray}

Due to the anomalous nature of the couplings, the loop integrals are in
general cutoff $\Lambda$ dependent. To calculate such dependence we use
dimensional regularization with a (modified) minimal subraction renormalization
scheme following the prescription in Ref.\cite{bur}.
Substituting $k' = k-(xp+yp')$ into eq.(3), the terms in odd powers of
$k'$ vanish. We find that among all the even power terms in $k'$, only
terms proportional to $g_1^Z$ and $g_5^Z$ will produce terms with no
powers in external momenta.
All other terms will be at least with  two  powers in
external momenta. Therefore their contributions to $K_L \rightarrow \mu^+\mu^-$
are suppressed by $m_d^2/M_W^2$, $m_s^2/M_W^2$ or $m_K^2/M_W^2$
compared with the contributions
from the $g_1^Z$ and $g_5^Z$ terms.
It is, then, obvious that the process $K_L \rightarrow
\mu^+\mu^-$ can only put useful constraints on $g_1^Z$ and $g_5^Z$ but not
the others. The $g_1^Z$ and $g_5^Z$ contributions to the effective $dsZ$
coupling is given by
\begin{equation}
H_{eff}(dsZ) = -{1\over 32\pi^2} g^3cos\theta_W V_{ls}V^*_{ld}F_A(x_l)
Z^\mu \bar d \gamma_\mu {1-\gamma_5\over 2}s + H.C.\;,
\end{equation}
where $x_l = m_l^2/M_W^2$
and the function $F_A(x)$ is given by
\begin{eqnarray}
F_A(x)&=&-g_1^Z{3\over 2}\Big( x \ln{\Lambda^2\over M_W^2}
+{x^2(2-x)\over (1-x)^2}\ln x +{11x-5x^2\over 6(1-x)}-{x\over 6}
\Big)\nonumber\\
&+&g_5^Z\Big( {3x\over 1-x} + {3x^2\ln x\over (1-x)^2})\;.
\end{eqnarray}

The amplitude for $K_L \rightarrow \mu^+\mu^-$ is obtained by exchanging
a virtual Z-boson between $d s$ and $\mu^+\mu^-$. At the quark level, we
obtain
\begin{equation}
H_{eff}= {G_F^2M_W^2\over 2\pi^2} cos^2\theta_W V_{ls}V^*_{ld}F_A(x_l)
\bar d\gamma^\mu{1-\gamma_5\over 2}s \bar \mu \gamma_\mu ({1-\gamma_5\over 2}
-2\sin^2\theta_W)\mu + H.C.\;.
\end{equation}

{}From this quark level effective Hamiltonian, we obtain the decay amplitude
\begin{equation}
M(K_L\rightarrow \mu^+\mu^-) = i{G_F^2M_W^2f_Km_\mu\over 2\sqrt{2}\pi^2}
Re(V_{ls}V^*_{ld}) cos^2\theta_WF_A(x_l)\bar\mu\gamma_5\mu\;.
\end{equation}
Here we have used: $<0|\bar s \gamma^\mu\gamma_5
d|K^0> = if_K p_K^\mu$, $p_K^\mu \bar \mu\gamma_\mu \mu = 0$,
and $p_K^\mu \bar \mu \gamma_\mu \gamma_5\mu = 2m_\mu\bar \mu \gamma_5\mu$.
We note that the vector current part does not contribute. For the same
reason the anomalous $WW\gamma$ interactions do not contribute to $K_L
\rightarrow \mu^+\mu^-$.

Combining the constribution from the MSM, we obtain the total amplitude
\begin{equation}
M^t(K_L\rightarrow \mu^+\mu^-) = i{G_F^2M_W^2f_Km_\mu\over 2\sqrt{2}\pi^2}
Re(V_{ls}V^*_{ld})\eta_lF(x_l)  \bar \mu\gamma_5
\mu\;,
\end{equation}
where $\eta_l$ are the QCD correction factors
which are of order one\cite{qcd}. The function
$F(x)$ is given by
\begin{equation}
F(x)=F_S(x)+cos^2\theta_WF_A(x)\;.
\end{equation}
with the MSM contribution $F_S(x)$ given by\cite{lim}
\begin{equation}
F_S(x) = -{2x\over 1-x} + {x^2\over 2(1-x)} - {3x^2\ln x\over 2(1-x)^2}\;.
\end{equation}

We are now ready to use experimental data to put constraint on $g_5^Z$.
The total branching ratio $Br^t$ for $K_L\rightarrow
\mu^+\mu^-$ is $(7.3\pm 0.4)
\times 10^{-9}$\cite{part}.
There are several different contributions to this decay which
can be
parametrized as $Br^t =R_{2\gamma} + R_{dis}$. Here $R_{2\gamma}$ is the
absorptive contribution due to two real photons in the intermediate state and
$R_{dis}$ is the dispersive contribution which contains the weak contribution
$R_W$ from eq.(9) and long distance contribution $R_{LD}$.
The absorptive part of the amplitude coming from
real photons in the intermediate state has been unambiguously determined
from the measured ratio
$Br(K_L\rightarrow \gamma\gamma) = (5.7\pm 0.27)\times 10^{-4}$\cite{part}.
This gives
$R_{2\gamma} = (6.83\pm 0.29)\times 10^{-9}$. The dispersive contribution
is then, $R_{dis} = (0.47\pm 0.56)\times 10^{-9}$. When extracting the weak
contribution from $R_{dis}$,
one faces the problem of subtracting the long distance
contribution. It has been
argued that this contribution is small compared
with the absorptive contribution by using
data from $K_L \rightarrow e^+ e^-\gamma$\cite{berg}.
The dispersive contribution may be solely due to weak contribution.
At the present the long distance contribution is not well
determined\cite{val}.
In our numerical analysis we will assume that $R_{dis}$ is saturated by the
weak contribution $R_W$.

To minimize uncertainties in $f_K$ we scale the rate $\Gamma(K_L\rightarrow
\mu^+\mu^-)_W$ due to the weak contribution
by $\Gamma(K^+\rightarrow \mu^+\nu_\mu)$. We have
\begin{eqnarray}
Br(K_L&\rightarrow& \mu^+\mu^-)={\tau^0\over \tau^+}Br(K^+\rightarrow
\mu^+\nu_\mu){\Gamma(K_L\rightarrow \mu^+\mu^-)_W\over\Gamma(K^+\rightarrow
\mu^+\nu_\mu)}\nonumber\\
&=&{\tau^0\over \tau^+}Br(K^+\rightarrow \mu^+\nu_\mu)
{G_F^2M_W^4\over 8\pi^4}{(1-4m_\mu^2/m_K^2)^{1/2}\over (1-m_\mu^2/m_K^2)^2}
{|Re(V_{sl}V^*_{dl}\eta_l F(x_l)|^2\over |V_us|^2}\;.
\end{eqnarray}
The branching ratio
$Br(K^+\rightarrow \mu^+\nu_\mu)$  is $63.5\% $,
and the lifetimes $\tau^0$ of $K_L$ and $\tau^+$ of $K^+$ are
$5.17\times 10^{-8} s$ and $1.237\times10^{-8} s$,
respectively\cite{part}.
We will use
$|V_{us}|=0.22$, and $\eta_l = 0.9$. The dominant contribution is from the
top quark in the loop. We must know the value for $Re(V_{ts}V^*_{td})$.
Unfortunately  this quantity is not well
determined at present. We will use the most recent estimate for
$|V_{td}|$ in Ref.\cite{bura} and take $Re(V_{ts}V^*_{td})$ to be in the range
$3.2\times 10^{-4}$ to $6.7\times 10^{-4}$.
In our analysis we will let the top quark
mass and the anomalous couplings $g_1^Z$ and $g_5^Z$ vary.

If $g_1^Z$ and $g_5^Z$ is set to zero, we obtain the MSM result.
Using the experimental data and
allowing the relevant KM matrix to span the allowed region, we find that
the top quark mass must be less than 240 GeV. This bound is weaker than
the bound from LEP data\cite{ting}.  In the following analysis,
we consider the cases  where one of  $g_1^Z$ and
$g_5^Z$ is not zero. In Tables 1, 2 and 3, we show the effects of non-zero
$g_1^Z$. Table 1. shows how $R_W/R_{2\gamma}$
varies with $g_1^Z$ for different cutoffs $\Lambda$.
We see that depending on the sign of $g_1^Z$,
the anomalous coupling $g_1^Z$ can either increase or decrease $R_W$.
Our results for the constraints on $g_1^Z$ at $2\sigma$ level for two
different cutoffs, $\Lambda = 1$ TeV and $\Lambda = 10$ TeV are shown in
Table 2. and 3. The constraints on $g_1^Z$ in Table 2. and 3. are for
$Re(V_{ts}V^*_{td})$ equal to  $3.2\times 10^{-4}$ and $6.7\times 10^{-4}$,
respectively. If $g_1^Z$ is positive the contribution from the anomalous
interaction has the same sign as the MSM contribution. $g_1^Z$ is
constrained to be in the range $-0.96$ to $0.57$ for $\Lambda = 1$ TeV.
The constraints
on $g_1^Z$ become tighter when the top quark mass is increased.
In Tables 4, 5, and 6, we show the effects of non-zero $g_5^Z$. This
contribution is cutoff independent. If $g_5^Z$ is positive, the contribution
has the opposite sign as that of the MSM.  $g_5^Z$ is constrained to be
between $-3.36$ to $5.67$.
Analysis with both $g_1^Z$ and $g_5^Z$ being non-zero can also be carried out.
In this case cancellations between the anomalous contributions may happen.
No significant additional
constraints on $g_1^Z$ and $g_5^Z$ can be obtained using
data only from $K_L \rightarrow \mu^+\mu^-$.

The same analysis can be carried out for $B\rightarrow \mu^+\mu^-$. In
this case the long distance contribution is expected to be small. When
experimental data for this decay will become available, one may obtain better
constraints on $g_1^Z$ and $g_5^Z$.

\acknowledgments
I would like to thank G. Valencia for useful discussions.
This work was supported in part by the Australian Research Council.

\begin{table}
\caption{$R_W/R_{2\gamma}$ vs. $g_1^Z$ for $m_t = 150 GeV$ and
$|Re(V_{ts}V_{td}^*)| = 5\times 10^{-4}$.}
\begin{tabular}{ccccccccccc}
$g_1^Z$&-1.0&-0.8&-0.6&-0.4&-0.2&-0.1&0.0&0.2&0.4&0.6\\ \hline
$\Lambda = 1$ TeV&&&&&&&&&&\\
$R_W/R_{2\gamma}$&2.25&1.27&0.57&0.15&$1.45\times
10^{-4}$
&0.03&0.13&0.54&1.22&2.18\\ \hline
$\Lambda = 10$ TeV&&&&&&&&&&\\
$R_W/R_{2\gamma}$&12.35&7.51&3.86&1.41&0.17&$7.3\times
10^{-4}$& 0.13&1.28&3.65&7.21
\end{tabular}
\label{table1}
\end{table}

\begin{table}
\caption{The constraints for $g_1^Z$ with $|Re(V_{ts}V_{td}^*)| = 3.2\times
10^{-4}$.}
\begin{tabular}{cccccc}
$m_t(GeV) $&100&125&150&175&200\\ \hline
$\Lambda = 1$ TeV&&&&&\\
$g_1^Z$&-0.99$\sim$ 0.59&-0.74$\sim$0.35&-0.59$\sim$
0.21&-0.51$\sim$0.12&-0.45$\sim$0.06\\ \hline
$\Lambda = 10$ TeV&&&&&\\
$g_1^Z$&-0.51$\sim$0.30&-0.37$\sim$0.17&-0.29
$\sim$0.10&-0.24$\sim$0.06&-0.20$\sim$0.03
\end{tabular}
\label{table2}
\end{table}

\begin{table}
\caption{The constraints for $g_1^Z$ with $|Re(V_{ts}V_{td}^*)| = 6.7
\times 10^{-4}$.}
\begin{tabular}{cccccc}
$m_t (GeV)$&100&125&150&175&200\\ \hline
$\Lambda = 1$ TeV&&&&&\\
$g_1^Z$
&-0.58$\sim$0.18&-0.45$\sim$0.06&-0.39$\sim$
$-9.6\times 10^{-4}$&-0.34$\sim$-0.04
&-0.31$\sim$-0.07
\\ \hline
$\Lambda = 10$ TeV&&&&&\\
$g_1^Z$&-0.30$\sim$0.09&-0.23$\sim$ 0.03
&-0.19$\sim$$-5.2\times 10^{-4}$&-0.16$\sim$-0.02
&-0.14$\sim$-0.03
\end{tabular}
\label{table3}
\end{table}
\newpage
\begin{table}
\caption{$R_W/R_{2\gamma}$ vs. $g_5^Z$ for $m_t = 150 GeV$ and
$|Re(V_{ts}V_{td}^*)|= 5\times 10^{-4}$.}
\begin{tabular}{cccccccccccc}
$g_5^Z$&6.0&5.0&4.0&3.0&2.0&1.37&1.0&0.0&-1.0&-2.0&-3.0\\ \hline
$R_W/R_{2\gamma}$&1.49&0.92&0.48&0.19&$2.8\times 10^{-2}$
&0.0&$9.4\time 10^{-3}$&
0.13&0.39&0.79&1.33
\end{tabular}
\label{table4}
\end{table}

\begin{table}

\caption{The constraints for $g_5^Z$ with $|Re(V_{ts}V^*_{td})|
= 3.2\times 10^{-4}$.}
\begin{tabular}{cccccc}
$m_t$(GeV) & 100 &125 &150 &175 &200\\ \hline
$g_5^Z$   & 5.67 $\sim$ -3.36&4.73$\sim$ -2.21& 4.2
$\sim$ -1.47
&3.91$\sim$ -0.94
&3.74$\sim$-0.53
\end{tabular}
\label{table5}
\end{table}

\begin{table}
\caption{The constraints for $g_5^Z$ with $|Re(V_{ts}V^*_{td})|
=6.7\times 10^{-4}$.}
\begin{tabular}{cccccc}
$m_t$(GeV) & 100 &125 &150 &175 &200\\ \hline
$g_5^Z$   &  3.32$\sim$ -1.00&2.92$\sim$ -0.40& 2.73$\sim$ 0.008
&2.64 $\sim$ 0.32
&2.63$\sim$0.59
\end{tabular}
\label{table6}
\end{table}
\end{document}